\documentclass[twocolumn,floatfix,aps,prd,showpacs]{revtex4}

\usepackage{graphicx}
\usepackage{bm}
\usepackage{epsf}

\newcommand{\beq}{\begin{equation}}
\newcommand{\eeq}{\end{equation}}
\newcommand{\beqn}{\begin{eqnarray}}
\newcommand{\eeqn}{\end{eqnarray}}
\newcommand{\pa}{\partial}

\begin{document}

\title{Nonaxisymmetric instability of rapidly rotating 
black hole in five dimensions}

\author{Masaru Shibata and Hirotaka Yoshino}

\affiliation{Yukawa Institute for Theoretical Physics, 
Kyoto University, Kyoto, 606-8502, Japan}

\affiliation{Department of Physics, University of Alberta, 
Edmonton, Alberta, Canada T6G 2G7}

\begin{abstract}
We present results from numerical solution of Einstein's equation 
in five dimensions describing evolution of rapidly rotating black
holes. We show, for the first time, that the rapidly rotating black
holes in higher dimensions are unstable against nonaxisymmetric
deformation; for the five-dimensional case, the critical value of spin
parameter for onset of the instability is $\approx 0.87$. 
\end{abstract}

\maketitle


\noindent
{\bf\em I Introduction}: Black holes (BHs) are the most strongly
self-gravitating objects in nature and also the simplest celestial objects, 
described by a small number of parameters.  Therefore, BHs are expected
to reflect the natures of gravitational theories in the 
clearest manner.  This fact motivates extensive studies of BHs not 
only for four-dimensional (4D) spacetime (see, e.g., \cite{Kip} for a historical
review) but also for higher-dimensional one (see, e.g., \cite{ER03} for a
review).

Since a possibility of BH formation in particle accelerators was
pointed out, studies for BHs in higher-dimensional spacetimes 
have been accelerated. If our space is a 3-brane in large \cite{ADD98} or
warped \cite{RS99} extra dimensions, the Planck energy could be of
$O({\rm TeV})$ that may be accessible with huge particle accelerators
like the Large Hadron Collider (LHC). In the presence of the extra
dimensions, mini BHs 
may be produced in the accelerators and its evidence may be detected.

A hypothetical phenomenology in the accelerators is as follows 
\cite{BHUA,reviews}:
In a high-energy particle
collision of a sufficiently small impact parameter, two particles 
merge to form a distorted rotating BH, and then, it perhaps relaxes to
a quasistationary state after emission of gravitational waves. Then,
the BH will evaporate by Hawking radiation 
due to a quantum-field-theory effect in a curved spacetime. 
The BH formation and subsequent evolution by gravitational radiation
can be described by classical general relativity \cite{GR04}. 
Since any approximation breaks
down for 
this phase because of its highly nonlinear nature,
numerical simulation in full general relativity is
the unique approach (see \cite{HEADON,SOY,SCPBHY,CP} for the 4D case).

One of the most important issues to be clarified is what type of a
black object is formed and whether it is stable or not. In
the 4D case, the formed object has to be a Kerr BH 
because of the uniqueness theorem of BHs
(e.g., \cite{unique} for a review), and the numerical analysis of a
Kerr BH strongly suggests that it should be stable
\cite{Kerr-stability}. These facts strongly constrain the possible
scenarios for the BH formation and subsequent evolution.  By contrast,
there is no uniqueness theorem in higher dimensions:
In the five-dimensional (5D) case, the black ring solution with 
the ringlike horizon is known in addition to 
the Myers-Perry BH solution with the spherical horizon 
(but see \cite{IM04} for uniqueness of
5D black holes with the spherical horizon topology).
Even if we assume that the black rings are unlikely to be
formed in two-particle systems, the scenario of mini BHs at accelerators
is still uncertain because there is
no proof for the stability of the higher-dimensional BHs \cite{ER03}.


Higher-dimensional BHs with certain parameters are known to be
unstable against axisymmetric perturbations.  Emparan and Myers
\cite{EM03} suggested that rotating BHs with a high spin parameter is
unstable for the spacetime dimensionality $D\ge 6$.  The reason is
that the rapidly rotating BHs have a high degree of ellipticity (i.e.,
the black membrane limit) and such objects are subject 
to the Gregory-Laflamme instability \cite{GL93}.  Very recently, Dias
{\it et al.} indeed showed, by a linear perturbation analysis, that
rapidly rotating BHs for $7\le D\le 9$ are unstable against
axisymmetric multiple-ring-like deformation \cite{Dias}. 

On the other hand, little is known for the stability of BHs against
the {\it nonaxisymmetric} perturbation, and also, for the
dimensionality $D=5$. Emparan and Myers~\cite{EM03} discussed the
possibility that the rapidly rotating BHs may be unstable also against
nonaxisymmetric perturbation for $D\ge 5$ using a thermodynamical
argument (i.e., by comparing the horizon area of a rotating BH and
that of two boosted Schwarzschild BHs with the same total energy and
angular momentum). However, the correspondence between the
thermodynamical and dynamical instabilities has not been well
established, and thus, a rigorous analysis is required in order to
clarify whether the rapidly rotating BHs are actually unstable against
nonaxisymmetric perturbation or not. 

In this paper, we tackle this stability issue of rapidly rotating 5D
BHs by fully solving Einstein's equation.
The merits of this approach are that a wide variety of instabilities
can be clarified with no ambiguity, and that the final fate after the
onset of the instabilities could be determined since the amplitude of
the perturbation from the background BH solution need not be assumed
to be small.  In this letter, we focus on the 5D BHs of only one spin
parameter because such a BH will be an outcome in the particle
accelerators. We shall explicitly show, for the first time, that
rapidly rotating 5D BHs are unstable against nonaxisymmetric
distortion.

\noindent
{\bf\em II Setting and Methodology}: We numerically study the
stability against nonaxisymmetric deformation of a 5D rotating BH of
single spin parameter. Its line element in the
Boyer-Lindquist-type coordinates is \cite{MP86}
\beqn
&&ds^2=-dt^2+{\mu \over \Sigma}(dt-a\sin^2\theta d\varphi)^2 
+{\Sigma \over \Delta} d\hat r^2 + \Sigma d\theta^2
\nonumber \\
&& ~~~~~~~+(\hat r^2 + a^2) \sin^2\theta d\varphi^2
+\hat r^2 \cos^2\theta d\chi^2,
\eeqn
where $\mu$ and $a$ are mass and spin parameters, respectively, and
$\Sigma:=\hat r^2 +a^2\cos^2\theta$ and $\Delta:=\hat r^2 +a^2 - \mu$. 
$\hat r=r_h:=(\mu-a^2)^{1/2}$ is the radius of event horizon. 
In five dimensions, $q:=|a|/\mu^{1/2}$ has to be less than the unity 
(i.e., $|a| < \mu^{1/2}$) by the requirement of global hyperbolicity. 
Note that we adopt the geometric units $G=1=c$ throughout this letter. 


We evolve this BH adopting the so-called quasi-isotropit coordinate
$r$ defined by $\hat r = r + r_h^2 /4r$. Here, the location of the
horizon is $r=r_h/2$.  With this transformation, the $t={\rm const.}$
hypersurface becomes spacelike everywhere for $ 0\leq r < \infty$, and
the physical singularity is not included in the initial surface
$t=0$. In this coordinate, the event horizon corresponds to the 
wormhole throat and, the sphere denoted by $r=0$ represents spacelike
infinity of another asymptotically flat region and becomes a
coordinate singularity.  However, the puncture approach of numerical
relativity (with the appropriate choice of the conformal factor and
gauge conditions) enables us to stably follow the evolution of BHs not
only for 4D spacetimes \cite{BHBH} but also for 5D ones \cite{YS09}.

Einstein's equation for a 5D vacuum spacetime are
solved in the higher-dimensional version of the BSSN
(Baumgarte-Shapiro-Shibata-Nakamura) formalism \cite{BSSN,YS09} with a
fourth-order finite differencing scheme in space and time. All the
equations are solved in the $4+1$ form with the Cartesian coordinates
$(x,y,z,w)$ where $w$ denotes the coordinate of the extra dimension. 
The following so-called puncture gauge conditions are adopted for the 
lapse function $\alpha$ and shift vector $\beta^i$, 
\beqn
&&\pa_t \alpha= -1.5 \alpha K,\\
&&\pa_t \beta^i=\eta_B B^i,~~
\pa_t B^i=\pa_t \tilde \Gamma^i - \mu^{-1/2} B^i,
\eeqn
where $B^i$ an auxiliary gauge variable, $K$ the trace part of the 
extrinsic curvature, $\tilde \Gamma^i$ is an auxiliary 
variable for the BSSN formalism, and we choose $\eta_B=1/3$. 

For this BH, rotational motion exists only in the $(x,y)$-plane. 
Because we consider the nonaxisymmetric stability against a bar 
deformation, the $z$- and $w$-axes directions are equivalent. 
Thus, a rotational Killing vector $(\pa/\pa \chi)^{\mu}$ is present
where $\chi=\tan^{-1}(w/z)$.  This fact motivates us to adopt a
cartoon method \cite{YS09,cartoon} that enables to solve 4+1
Einstein's evolution equation in the 3D grid $(x,y,z)$ 
(see \cite{YS09}).

Numerical simulation was performed with two codes. One is the code
reported in \cite{YS09}. This code prepares a nonuniform grid for
which the region around the BH is resolved with a high accuracy and
the distant region is covered by a sparse grid spacing. The other is
the {\small SACRA5D} code in which an adaptive mesh refinement algorithm
is implemented. This is the code extended from the {\small SACRA} code
which was originally developed for simulations of 4D spacetimes
\cite{SACRA}. {\small SACRA5D} has been tested by solving the problems
listed in \cite{YS09}. Several simulations were performed for both
codes and we have confirmed that the two codes provide the same 
conclusion concerning the stability of BHs.

Although the numerical solutions show a behavior of convergence with
improving grid resolution, the convergence speed depends strongly on
the spin parameter, $q$. 
In this work, we measured the accuracy by monitoring the area of
the apparent horizon by evolving non-perturbed rotating BHs and 
determined the required resolution: We evolved the BHs at least for
$50\mu^{1/2}$ and checked required resolution with which the horizon
area remains approximately constant within $1\%$ error.  For $q \alt
0.7$, we found that the grid spacing near horizon $\Delta x \approx
0.03\mu^{1/2}$ is small enough for this requirement. However, for $q
\agt 0.8$, the required value for $\Delta x$ changes steeply: For
$q=0.8$ with $\Delta x/\mu^{1/2}=0.015$ and $0.0225$, the errors
at $t=50\mu^{1/2}$ are 0.2\% and 1.2\%, respectively.  
For $q=0.85$ and 0.88, $\Delta x/\mu^{1/2}$ should be smaller 
than $\approx 0.01$ and $0.008$,
respectively, to guarantee the error within 1\% at $t=50\mu^{1/2}$.
For $q \geq 0.89$, accurate simulation up to $t=50\mu^{1/2}$ is not
feasible even for $\Delta x=0.008 \mu^{1/2}$. Therefore, for this case, the
simulations were performed up to the time at which numerical error of
apparent horizon area exceeds 3\% error (at $t < 50 \mu^{1/2}$) choosing
$\Delta x=0.008 \mu^{1/2}$. Even these short-term simulations are long 
enough to show that rapidly rotating BHs are unstable. 

To investigate the nonaxisymmetric stability, we initially superimpose 
a small nonaxisymmetric perturbation on the rotating BH solution. 
Specifically, we perturb a conformal factor of the 4D space defined by 
$W={\rm det}(\gamma_{ij})^{-1/4}$ ($\gamma_{ij}$ is the 4D
space metric) as
\beq
W=W_0 [1 + A\mu^{-1}(x^2 - y^2) \exp(-r^2/2r_h^2)],
\eeq 
where $W_0$ is the nonperturbed solution. In the following, 
we choose $A=0.005$.  Simulations were also done with values 
$A=0.01$ and $0.02$, but the evolution of the perturbed
part obeys a scaling relation (e.g., $W/(W_0A)$ behaves approximately
in the same manner). Thus, the magnitude of $A$ does not change the
conclusion in this paper, as far as $A \ll 1$.

During numerical simulation, we monitor two quantities for determining 
the stability against the nonaxisymmetric deformation.  One is a
deformation parameter of BH horizon. To define this parameter, we 
calculate the circumferential radii of the apparent horizon along several 
meridians. Specifically, we measure the proper length of the meridians
(referred to as $l_{\varphi}$) 
for $\varphi=0$ (and $\pi$), $\pi/4$ (and $5\pi/4$), $\pi/2$ (and
$3\pi/2$), and $3\pi/4$ (and $7\pi/4$), and then, 
define the deformation parameter
\beq
\eta=\sqrt{(l_0-l_{\pi/2})^2+(l_{\pi/4}-l_{3\pi/4})^2}/l_0
\eeq
The other is the gravitational waveform in a wave zone 
defined along the $z$-axis by 
\beq
h_+ \equiv r^{3/2} \mu^{-3/4} (\tilde \gamma_{xx}-\tilde \gamma_{yy})/2,
\eeq
where $\tilde \gamma_{ij}$ is the conformal 4D metric. 
This quantity is regarded as the $+$ mode of gravitational waves. 

In the 4D case, the amplitude of these quantities decreases 
exponentially with time as far as $q < 1$, because the BHs are stable; 
the damping rate is primarily determined by the fundamental 
quasinormal modes.  However, this is not the case in the 5D case 
(see below). 

\noindent
{\bf\em III Results}: Figure 1 plots the evolution of $\eta$ as a
function of time for $q=0.8$--0.89.  Irrespective of the values of
$q$, $\eta$ quickly increases for $t \alt 10 \mu^{1/2}$.  This is
simply because the perturbation initially given relaxes.  Indeed, the
value of $\eta$ after this relaxation phase is of $O(10^{-3})$, which
is the same order as the initial perturbation amplitude.  However, the
evolution after the relaxation depends strongly on the values of $q$.
For $q \leq 0.86$, the value of $\eta$ decreases; in particular for $q
\leq 0.80$, this damping occurs very quickly. By contrast, for $q \agt
0.88$, the perturbation grows exponentially for $t/\mu^{1/2} \agt 25$.
The growth rate does not depend on the initial perturbation amplitude.
Thus, the BHs of $q \agt 0.88$ are dynamically unstable against the
nonaxisymmetric deformation. For $q=0.87$, the value of $\eta$ grows
slowly with time. This suggests that this value is near the critical
value for the onset of this instability.

\begin{figure}[t]
\vspace{-5mm}
\epsfxsize=2.8in
\leavevmode
\epsffile{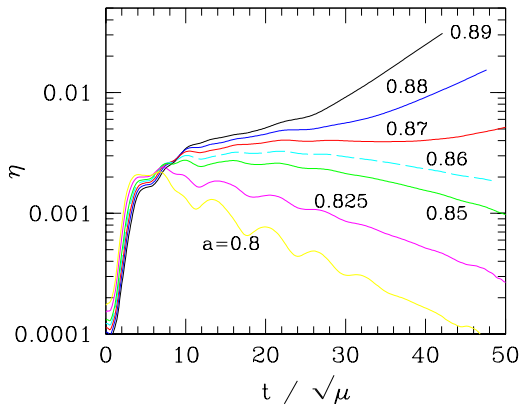}
\vspace{-15mm}
\caption{Evolution of deformation parameter $\eta$ for
$a/\mu^{1/2}=0.80$--0.89.
\label{FIG1}}
\end{figure}

\begin{figure}[t]
\vspace{-5mm}
\epsfxsize=2.8in
\leavevmode
\epsffile{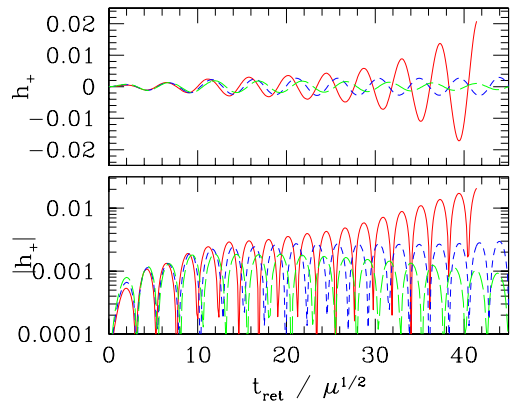}
\vspace{-15mm}
\caption{$h_+$ and its absolute value as functions of retarded time 
for  $a/\mu^{1/2}=0.85$, 0.87, and 0.89 (dashed, long-dashed, and 
solid curves). $h_+$ is extracted for $r \sim \lambda$ where $\lambda$ 
is a gravitational-wave length. 
\label{FIG2}}
\end{figure}

Figure 2 plots $h_+$ and $|h_+|$ as a function of a retarded time
$t_{\rm ret} \equiv t-r$. The amplitude for $t \agt 25\mu^{1/2}$
exponentially increases with time for $q=0.89$ as in Fig.~1. 

From gravitational waveforms, it is possible to extract the frequency
of gravitational waves, $f$. Figure 3 plots a characteristic frequency
of gravitational waves as a function of $q$. Here, the characteristic
frequency is determined by performing the Fourier transformation of
$h_+(t)$ and then by identifying its peak. Figure 3 shows that $f$
increases with $q$ for $q \geq 0.8$.  This property agrees with that
of the fundamental quasinormal mode for 4D Kerr BHs
\cite{L85}. Furthermore, the frequency determined for $a=0$ gives $f
\approx 0.15/\mu^{1/2}$ which agrees with that obtained by a
perturbation analysis \cite{pert5d}. Hence, it is natural to consider
that emitted gravitational waves are associated with the quasinormal
modes of the BHs.

A remarkable fact is that the angular velocity defined by $\omega
:= \pi f$ is smaller than that of BHs $\Omega=a/(r_h^2 + 
a^2)=a/\mu$ for a large value of $a$. Remember that the total radiated
energy ($\Delta E$) and angular momentum ($\Delta J$) approximately 
obey a relation $\omega \Delta J = \Delta E$, 
and the first law of the BH thermodynamics allows 
us to obtain the variation in the BH area $\delta A$ as $\kappa \delta
A/8\pi = \Omega \delta J - \delta E$ where $\kappa$ is the surface
gravity of the BH horizon, and $\delta E$ and $\delta J$ are the
variation of the energy and angular momentum of the BH. If $\delta E$
and $\delta J$ are equal to $\Delta E$ and $\Delta J$, respectively,
we obtain the relation
\beqn
\kappa\delta A= 8\pi (\Omega - \omega)\delta J. 
\eeqn
For $\Omega > \omega$, $\delta A$ becomes positive, and thus, the
evolution by emission of gravitational waves is allowed: A rapidly
rotating BH in five dimensions may be unstable against gravitational
radiation reaction (although this is not a sufficient condition). This
fact seems to be at least a part of the reason why rapidly rotating
BHs are unstable against nonaxisymmetric deformation.

\begin{figure}[t]
\vspace{-3mm}
\epsfxsize=2.7in
\leavevmode
\epsffile{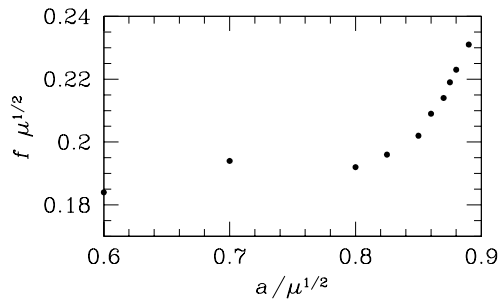}
\vspace{-26mm}
\caption{Characteristic frequency of gravitational waves as 
a function of $a/\mu^{1/2}$. 
\label{FIG3}}
\end{figure}



\noindent
{\bf\em IV Summary}: We showed, for the fist time, that rapidly
rotating BHs in 5D vacuum spacetime are unstable against
nonaxisymmetric deformation. The critical value for the spin parameter
is $\approx 0.87$.  The critical value we find is close to the value
predicted by Emparan and Myers, $\approx 0.85$ \cite{EM03}.  This
suggests that their heuristic thermodynamic argument concerning the
stability of higher-dimensional rotating BHs may be reliable.

Unfortunately, the present numerical simulation cannot clarify the final
fate of the unstable BHs, because it is not easy to maintain the
numerical accuracy to follow the unstable BH for a sufficiently long
time. To clarify the final fate, a simulation with a much better grid
resolution is required.  This issue is left for the future study. 

There will be at least two possible fates for an unstable BH.  One is
that the perturbation grows until the unstable BH fragments into two
BHs \cite{EM03}, and the other is that the growth of the perturbation
saturates at a stage when the emission rate of gravitational waves is
large enough to quickly carry angular momentum of the BHs for
stabilization. This issue is quite similar to the nonaxisymmetric
dynamical stability of rotating stars in four dimensions: For many
cases, rotating stars are dynamically unstable if the ratio of
rotational kinetic energy to gravitational potential energy is larger
than $0.27$ or the ratio of the polar axial length to the equatorial
axial length is smaller than $\sim 0.2$ \cite{CH69}.  This condition
holds irrespective of equations of state, as far as the degree of
differential rotation is not extremely large (e.g., \cite{SKE} and
references therein).  Dynamically unstable rotating stars evolve after
the onset of the instability via several mechanisms such as angular
momentum transfer and gravitational radiation reaction. Then, they
settle to a new stable state. A noteworthy fact is that fragmentation 
rarely occurs for stars (for tori, it may occur). 

For the 5D BHs, the ratio of the meridional circumferential length to
the equatorial one decreases with $q$, and $C_m/C_e=0.38$ for
$q=0.87$. Thus, a BH is unstable for $C_m/C_e \alt 0.38$. For the 4D
case, the minimum value of this ratio is at most $C_m/C_e=0.64$ (for
the extreme Kerr BH); all the BHs are not very oblate.  This may be
the reason that BHs are stable in four dimensions.  These facts
suggest that a sufficiently oblate BH with $C_m/C_e \alt 0.4$ is
dynamically unstable against nonaxisymmetric deformation irrespective
of the dimensionality, as in rapidly rotating stars.  Rapidly rotating
BHs in any higher dimensions can have small values of $C_m/C_e$
\cite{ER03}, and thus, this issue should be also explored by
numerical-relativity simulation. 


\noindent
{\bf\em Acknowledgments}: We thank T. Tanaka and T. Shiromizu for
helpful discussions. Numerical computations were in part performed on
the NEC-SX9 at CfCA in NAO of Japan and on the NEC-SX8 at Yukawa
Institute for Theoretical Physics in Kyoto University. This work was in
part supported by Grant-in-Aid for Scientific Research (21340051) and
by Grant-in-Aid for Scientific Research on Innovative Area (20105004)
of the Japanese Monbukagakusho.  HY is supported by JSPS (program of
Postdoctoral Fellow for Research Abroad).


\end{document}